\documentclass[aps, prd, twocolumn, groupedaddress ]{revtex4-1}
\usepackage{amsmath,amssymb}
\usepackage{graphicx}
\usepackage{color}
\usepackage{slashed} 
\usepackage{hyperref}
\begin{document}
\newcommand{\nn}{\nonumber}
\def\d{{\mathrm{d}}}
\def\lint{\hbox{\Large $\displaystyle\int$}}   
\def\hint{\hbox{\huge $\displaystyle\int$}}  
\title{\bf\Large Special-case closed form of the Baker--Campbell--Hausdorff formula}
\author{Alexander Van--Brunt\,}
\email[]{alexandervanbrunt@gmail.com}
\author{Matt Visser\,}
\email[]{matt.visser@msor.vuw.ac.nz}
\affiliation{ \mbox{School of Mathematics, Statistics, and Operations Research,}
Victoria University of Wellington; \\
PO Box 600, Wellington 6140, New Zealand.\\}
\date{12 January 2015; 13 April 2015; \LaTeX-ed \today}
\begin{abstract}
The Baker--Campbell--Hausdorff formula is a general result for the quantity $Z(X,Y)=\ln( e^X e^Y )$, where $X$ and $Y$ are not necessarily commuting.  For  completely general commutation relations between $X$ and $Y$, (the free Lie algebra), the general result is somewhat unwieldy. However in specific physics applications the commutator $[X,Y]$, while non-zero, might often be relatively simple, which sometimes leads to explicit closed form results. We consider the special case $[X,Y] = u X + vY + cI$,  and show that in this case the general result reduces to
\[
Z(X,Y)=\ln( e^X e^Y ) = X+Y+ f(u,v) \; [X,Y].
\]
Furthermore we explicitly evaluate the symmetric function $f(u,v)=f(v,u)$, demonstrating that
\[
f(u,v) = {(u-v)e^{u+v}-(ue^u-ve^v)\over u v (e^u - e^v)},
\]
and relate this to previously known results. For instance this result includes, but is considerably more general than, results obtained from either the Heisenberg commutator $[P,Q]=-i\hbar I$ or the creation-destruction commutator $[a,a^\dagger]=I$.

\end{abstract}
\maketitle

\section{Introduction}

Various partial results leading to what is now called the  Baker--Campbell--Hausdorff formula have by now been in circulation for well over 100 years. 
A recent study of the early history can be found in~\cite{early-history}.  The basic question being addressed is this: What can one say about the quantity  $Z(X,Y)=\ln( e^X e^Y )$ whenever $X$ and $Y$ do not commute? For concrete examples of this phenomenon one could think of matrices and/or linear operators, but the context could be as general as an abstract free Lie algebra. See for instance references~\cite{Dynkin1, Dynkin2, Goldberg, Sack, Wilcox, Newman-Thompson, Reinsch, Casas-Murua}. 
Perhaps the most commonly quoted result is this:
\begin{equation}
Z(X,Y)=\ln( e^X e^Y ) = X + Y + {1\over2}\;[X,Y] + \dots
\end{equation}
Less commonly, a few more explicit terms are added:
\begin{eqnarray}
Z(X,Y)&=&\ln( e^X e^Y ) = X + Y + {1\over2}\;[X,Y]
\nonumber\\
&& + {1\over12} \Big( [X,[X,Y]] - [Y,[X,Y]] \Big)
\nonumber\\
&& - {1\over24} [Y,[X,[X,Y]]] + \dots
\end{eqnarray}
Unfortunately the expansion rapidly becomes extremely unwieldy, with the (average) number of terms growing rapidly with the level of commutators being retained~\cite{Dynkin1, Dynkin2}. Even though explicit computer-aided computations can easily be carried out to 10 or even more nested commutators, the resulting formulae are simply too cumbersome to be usefully written down on paper.

\section{General Commutators} 

In contrast, using the quite common notation $L_A B = [A,B]$, the exact fully general result can be written in the quite standard form~\cite{early-history}
\begin{eqnarray}
\label{BCH}
Z(X,Y)&=&\ln( e^X e^Y ) = X + Y 
\nonumber\\
&& - \int_0^1 \d t  \sum_{n=1}^\infty {(I-e^{L_X} e^{tL_Y})^n\over n(n+1)} \; Y.
\end{eqnarray}
Expanding this sum rapidly becomes quite complicated.  
Using $e^{tL_Y} Y = Y$, it is quite useful to rewrite this general formula as 
\begin{eqnarray}
Z(X,Y)&=&\ln( e^X e^Y ) = X + Y 
\\
&& - \int_0^1 \d t  \sum_{n=1}^\infty {(I-e^{L_X} e^{tL_Y})^{n-1} \over n(n+1)} \; (I - e^{L_X}) Y.
\nonumber
\end{eqnarray}
The advantage of doing this is that one now has
\begin{eqnarray}
&&Z(X,Y)=\ln( e^X e^Y ) = X + Y 
\\
&&\quad  + \int_0^1 \d t  \sum_{n=1}^\infty {(I-e^{L_X} e^{tL_Y})^{n-1} \over n(n+1)} \; {( e^{L_X}-I)\over L_X} \; [X, Y].
\nonumber
\end{eqnarray}
This focuses attention on the fact that the Baker--Campbell--Hausdorff formula can be expressed in terms of nested commutators acting on the elementary commutator $[X,Y]$. Thus if we have some extra information regarding the elementary commutator $[X,Y]$, then there is some hope that the Baker--Campbell--Hausdorff result might simplify. Similar (but distinct) formulae can be extracted directly from Dynkin's expansion~\cite{Dynkin1, Dynkin2}. 

\section{Two Very Special Commutators} 

It is quite common in physically relevant  situations to have $[X,Y]=cI$. (This occurs for instance for both the Heisenberg commutator $[P,Q]=-i\hbar I$, and the creation-destruction commutator $[a,a^\dagger]=I$). In that specific case $L_X[X,Y]=0=L_Y[X,Y]$, and the entire Baker--Campbell--Hausdorff series collapses down to just the $n=1$ term. We then have the exact result
\begin{equation}
Z(X,Y)=\ln( e^X e^Y ) \to X + Y + {1\over2}\;[X,Y].
\end{equation}
We use the symbol $\to$ in the sense of ``simplifies to'', under the stated conditions on the commutator $[X,Y]$.
Another well-known result, valid whenever $[X,Y] = v Y$, is that
\begin{eqnarray}
Z(X,Y)&=&\ln( e^X e^Y ) \to   X + {v Y\over 1-e^{-v}} 
\\
&=& X + Y + {ve^v-e^v+1\over e^v-1}\; Y 
\\
&=& X + Y + {ve^v-e^v+1\over v(e^v-1)}\; [X,Y].
\end{eqnarray}
For example: The commutator $[X,Y] = v Y$ implies that $X$ acts as a ``shift operator'', (a ``ladder operator''), for $Y$, thus allowing one to invoke the techniques of Sack~\cite{Sack}. 
This same result can also be extracted from equation (7.9) of Wilcox~\cite{Wilcox}; but only after some manipulations. More prosaically, we note that $[X,Y] = v Y$ implies that $L_Y X = -v Y$, and so $L_Y^m X = 0$ for $n\geq 2$. Thus for this specific commutator $e^{tL_Y} X \to (I+tL_Y) X$, so that the BCH series of equation (\ref{BCH}) simplifies to something that can, (with a little work), be explicitly integrated and summed.
We shall now generalize these results somewhat further.

\section{Special Commutator} 

Consider the special-case commutator
\begin{equation}
[X,Y] = u X + v Y + c I.
\end{equation}
This is already considerably more general than the two very special cases mentioned above, but still tractable enough to be interesting. 
For this commutator we have
\begin{equation}
L_X [X,Y] = v [X,Y]; \qquad L_Y [X,Y] = - u [X,Y].
\end{equation}
(Note that $c$ has dropped out of these formulae.)
This means that in the Baker--Campbell--Hausdorff series the nested commutators all collapse as follows:
\begin{equation}
e^{L_X}[X,Y] \to e^v [X,Y]; \qquad e^{tL_Y}[X,Y] \to e^{-tu} [X,Y].
\end{equation}
Therefore
\begin{eqnarray}
&&Z(X,Y)=\ln( e^X e^Y ) \to X + Y 
\\
&&\quad  + \int_0^1 \d t  \sum_{n=1}^\infty {(1-e^{v} e^{-tu})^{n-1} \over n(n+1)} \; {( e^{v}-1)\over v} \; [X, Y].
\nonumber
\end{eqnarray}
(Note that $c$ has dropped out of this formula also.)
This is enough to guarantee that in this situation
\begin{equation}
Z(X,Y)=\ln( e^X e^Y ) \to X+Y+ f(u,v) \; [X,Y],
\end{equation}
where $f(u,v)$ is some function still to be determined. 

First, we note that the function $f(u,v)=f(v,u)$ is symmetric. This can be established as follows.
Since 
\begin{equation}
(e^X e^Y)^{-1}= e^{-Y} e^{-X}, 
\end{equation}
we know that 
\begin{equation}
Z(-Y,-X)=-Z(X,Y).
\end{equation} 
By reversing our special commutator we see
\begin{equation}
[-Y,-X] =  v(-Y) +u(-X)- cI.
\end{equation}
Furthermore
\begin{equation}
L_{-Y} [-Y,-X] = u [-Y,-X],
\end{equation}
and
\begin{equation}
L_{-X} [-Y,-X] = - v [-Y,-X],
\end{equation}
are nested commutators with the roles of $u\leftrightarrow v$ interchanged. 
Combining these facts now leads to the desired symmetry:
\begin{eqnarray}
&&X+Y+ f(u,v) \; [X,Y] =Z(X,Y) = - Z(-Y,-X) 
\nonumber\\
&& \quad = - \{ (-X)+(-Y) + f(v,u) [-Y,-X]\}
\nonumber\\
&& \quad = X+Y + f(v,u) [X,Y].
\end{eqnarray}
Secondly, we note the explicit result
\begin{equation}
f(u,v) =  {( e^{v}-1)\over v}  \int_0^1 \d t  \sum_{n=1}^\infty {(1-e^{v} e^{-tu})^{n-1} \over n(n+1)}.
\end{equation}
The sum and integral are easily carried out, with the result that
\begin{equation}
f(u,v) = {(u-v)e^{u+v}-(ue^u-ve^v)\over u v (e^u - e^v)}.
\end{equation}
A key step is to note that we have the Taylor series
\begin{equation}
{\ln(x) x \over x-1}  = 1 - \sum_{n=1}^\infty {(1-x)^n\over n(n+1)}.
 \end{equation}
One might also wish to check the cases $u=0$, $v=0$, and $u=v$ explicitly. Also note that the $f(u,v)$ above does in fact exhibit the desired symmetry, even if this is not obvious before one performs the sum and integral.

Sometimes it is more useful to cast this result as
\begin{equation}
f(u,v) = {u e^u(e^v-1) - v e^v (e^u-1)\over u v (e^u - e^v)},
\end{equation}
or even
\begin{equation}
f(u,v) = {u (1-e^{-v}) - v (1-e^{-u})\over u v (e^{-v} - e^{-u})}.
\end{equation}
Applying the l'Hospital rule, it is now easy to check that
\begin{equation}
f(0,0) = {1\over2}
\end{equation}
as it should. Furthermore
\begin{equation}
f(0,v) = {ve^v-e^v+1\over v(e^v-1)} 
\end{equation}
as it also should.  This can also be re-cast in a somewhat more symmetrical form as
\begin{equation}
f(0,v)= {1\over2} +{1\over2} \coth(v/2) - {1\over v}.
\end{equation}
Along the diagonal we have
\begin{equation}
f(u,u) = {e^u-1-u\over u^2} = {1\over2} +  {e^u-(1+u+{1\over2} u^2)\over u^2}.
\end{equation}
Along the anti-diagonal we have
\begin{equation}
f(u,-u) = {\tanh(u/2)\over u}.
\end{equation}
Overall, the form of the function $f(u,v)$, while quite tractable, is not something that would have been easy to guess from first principles.
(We have also verified our result for $f(u,v)$ via an independent brute force computation directly from Dynkin's formula.)

In view of the fact that
\begin{equation}
[sX,tY] = st(uX+vY+cI) = ut(sX) +  sv(tY)+stcI,
\end{equation}
we see that
\begin{equation}
Z(sX,tY) = \ln(e^{sX} e^{tY}) = sX+tY + st f(ut,sv) [X,Y].
\end{equation}
Though superficially more general, this result is in fact implicit in our previous result.

Finally, note that we can replace the $cI$ terms in the commutator with $cE$, where $E$ is any object that commutes with both $X$ and $Y$, (that is $L_X E = 0 = L_Y E$), without needing to change any of the discussion above.

\section{A Shifted BCH formula} 

Note that our commutator
\begin{equation}
[X,Y] = u X + v Y + c I
\end{equation}
can be re-written as
\begin{equation}
\left[X, {u\over v} X + Y + {c\over v} I \right] =  v\left( {u\over v} X + Y + {c\over v} I \right).
\end{equation}
Defining $\tilde Y= {u\over v} X + Y + {c\over v} I $ this reduces to $[X,\tilde Y]=v\tilde Y$, which allows us to assert
\begin{equation}
Z(X,\tilde Y)=\ln( e^X e^{\tilde Y} ) \to   X + {v \tilde Y\over 1-e^{-v}}.
\end{equation}
This now implies
\begin{equation}
\ln\left(e^X e^{(u/v)X+Y}\right) =  X + {u X +v  Y\over 1-e^{-v}} +{c \, I \, (e^{-v} -1 + v) \over v(1-e^{-v})}.
\end{equation}
While this is a perfectly correct relation, it unfortunately not of the form we were aiming for --- the LHS is not of the form $\ln\left(e^X e^{Y}\right)$, and there is no simple trick to convert the LHS to that form. Similar issues arise in the Wilcox article~\cite{Wilcox}; while that article provides many formulae extracted via parameter differentiation, none of those formulae are of the same form as the key result of the present article.

\section{$2\times2$ matrix representation} 

Once one has seen the result derived directly from a specific instance of the Baker--Campbell--Hausdorff series it is relatively easy to then check it using a specific $2\times2$ matrix representation. Consider the two (craftily chosen) matrices
\begin{equation}
X = \left[\begin{array}{cc}{v\over2} & 1\\ 0 & -{v\over2}\end{array}\right]; 
\qquad
Y = \left[\begin{array}{cc}-{u\over2} & 1\\ 0 & {u\over2}\end{array}\right].
\end{equation}
Then it is easy to check that
\begin{equation}
[X,Y] =  \left[\begin{array}{cc}0 & u+v\\ 0 & 0\end{array}\right] = u X + v Y.
\end{equation}
Thus these two simple $2\times2$ matrices provide us with an explicit representation of the $c=0$ sub-case of our special commutator.
(Generalizing to $c\neq0$ is straightforward but tricky, see below.)
It is now easy (eg, via {\sf Maple} or some equivalent) to calculate 
\begin{equation}
\exp X =\left[\begin{array}{cc}e^{v/2} & {\cosh(v/2)\over v/2} \\ 0 &e^{-{v/2}}\end{array}\right]; 
\end{equation}
and
\begin{equation}
\exp Y = \left[\begin{array}{cc}e^{-{u/2}} &  {\cosh(u/2)\over u/2}\\ 0 & e^{u/2}\end{array}\right].
\end{equation}
Brute force calculation of the $2\times 2$ matrix logarithm (eg, via {\sf Maple} or some equivalent) yields
\begin{equation}
\ln\left( \exp X \exp Y\right) = \left[\begin{array}{cc} {v-u\over2}&  2 + (u+v) \, f(u,v)\\ 0 & {u-v\over2}\end{array}\right],
\end{equation}
with \emph{exactly the same} function $f(u,v)$ as we previously encountered. That is, for these specific $2\times2$ matrices we have
\begin{equation}
\ln\left( \exp X \exp Y\right) = X + Y + f(u,v)\; [X,Y].
\end{equation}
This is certainly a consistency check on our key result, but it is actually much more than that. 

To now deal with the situation where $c\neq 0$,  let $p$ be arbitrary,  and set
\begin{equation}
X = - {c\,p\over u}\; I + \left[\begin{array}{cc}{v\over2} & 1\\ 0 & -{v\over2}\end{array}\right]; 
\quad
Y = - {c(1-p)\over v}\; I +\left[\begin{array}{cc}-{u\over2} & 1\\ 0 & {u\over2}\end{array}\right].
\end{equation}
Then for all values of $p$ we have
\begin{equation}
[X,Y] =  \left[\begin{array}{cc}0 & u+v\\ 0 & 0\end{array}\right] = u X + v Y + cI.
\end{equation}
The rest of the computation carries through as before, again
with \emph{exactly the same} function $f(u,v)$ as previously encountered. 

Note that even though this particular computation is based on a specific $2\times 2$ matrix representation of our special commutation relation, the only feature which the computation actually depends on is the existence of that special commutation relation. That is, once one thinks about it more carefully, this computation actually provides an independent proof of our desired result. 

\section{Braiding relations} 

Let us now apply the special commutation relation
\begin{equation}
[X,Y] = u X + v Y + c I,
\end{equation}
to the general Baker--Hausdorff lemma
\begin{equation}
e^X Y e^{-X} = \exp(L_X) Y = \sum_{n=0}^\infty {L_X^n\over n!}\; Y,
\end{equation}
and to the general braiding relation
\begin{eqnarray}
e^X e^Y &=& e^X e^Y (e^{-X} e^X) =   (e^X e^Y e^{-X} ) e^X 
\nonumber\\
&=& e^{e^X Y e^{-X}} e^X = e^{\exp(L_X) Y} e^X.
\end{eqnarray}
We split off the first term in the expansion and note
\begin{equation}
e^X Y e^{-X}  = Y +  \sum_{n=1}^\infty {L_X^{n-1}\over n!}\; [X,Y].
\end{equation}
In view of the fact that for our special commutator we have $L_X [X,Y]=-v[X,Y]$, the sum collapses to
\begin{equation}
e^X Y e^{-X}  \to Y +  \sum_{n=1}^\infty {(-v)^{n-1}\over n!}\; [X,Y],
\end{equation}
that is
\begin{equation}
e^X Y e^{-X}  \to  Y + {1-e^{-v}\over v}\;[X,Y].
\end{equation}
The braiding relation thus specializes to
\begin{equation}
e^X e^Y = e^{Y + v^{-1} (1- e^{-v})\;[X,Y]} \;e^X.
\end{equation}
In a completely analogous manner we have
\begin{equation}
e^Y X e^{-Y}  \to  X + {e^{u}-1\over u} \;[X,Y],
\end{equation}
and
\begin{equation}
e^Y e^X = e^{X + u^{-1} (e^{u}-1)\;[X,Y]} \;e^Y.
\end{equation}

\section{Application to squeezed states} 

This formalism also leads to some interesting results for squeezed states~\cite{Fisher, Schumaker, Truax, Shanta:1993, Nieto:1993, Nieto:1996}. 
Begin by considering the usual creation-destruction algebra
\begin{equation}
[a,a^\dagger]=I.
\end{equation}
It is easy to check that
\begin{equation}
[a^2,(a^\dagger)^2] =  4 \left(a^\dagger a\right) + 2 I = 4 N +2 I;
\end{equation}
while
\begin{equation}
[a^2, (a^\dagger a) ] = [a^2,N] =2 a^2;
\end{equation}
and
\begin{equation}
[(a^\dagger)^2, (a^\dagger a) ] = [(a^\dagger)^2, N ]  = -2 (a^\dagger)^2.
\end{equation}
Now these last two formulae are specific instances of our special commutator, which is enough to imply the nontrivial (and perhaps unexpected) results
\begin{eqnarray}
Z\left(sa^2,tN\right) &=& \ln\left(e^{sa^2} e^{t N}\right)
\nonumber\\
&=& s a^2 + t N + 2st \,f(2t,0) a^2
\nonumber\\
&=& {2ste^{2t}\over e^{2t}-1} \,a^2 + t N;
\end{eqnarray}
and
\begin{eqnarray}
Z\left(s(a^\dagger)^2,tN\right) &=&  \ln\left(e^{s(a^\dagger)^2} e^{t N}\right) 
\nonumber \\
&=& s (a^\dagger)^2 + t N - 2 s t \,f(-2t,0) (a^\dagger)^2
\nonumber\\
&=& -{2ste^{-2t}\over e^{-2t}-1} \,(a^\dagger)^2 + t N.
\qquad
\end{eqnarray}
Furthermore, let us now consider the somewhat more complicated commutator
\begin{eqnarray}
&&[|w|N+\bar w a^2,|w|N+w(a^\dagger)^2] 
\nonumber\\
&&\qquad = 2|w|\bar w a^2 +2 |w| w  (a^\dagger)^2 +w\bar w (4N+2I),
\qquad
\end{eqnarray}
where $w$ is an arbitrary complex number.
Rewriting this as
\begin{eqnarray}
&&[|w|N+\bar w a^2,|w|N+w(a^\dagger)^2] 
\nonumber\\
&&\qquad
= 2|w|(|w| N+\bar w a^2) +2|w| (|w| N +w  (a^\dagger)^2) 
\nonumber\\
&&\qquad\qquad
+ 2 |w|^2 I,\qquad
\end{eqnarray}
we recognize another specific instance of our special commutator, (now with $u=2|w|=v$, and $c=2|w|^2$). 

Consequently we have
\begin{eqnarray}
&&Z\left(|w|N+\bar w a^2,|w|N+w(a^\dagger)^2\right) 
\nonumber\\
&&\quad= 2|w|N+\bar w a^2+w(a^\dagger)^2
\nonumber\\
&&\qquad+ f(2|w|,2|w|) [|w|N+\bar w a^2,|w|N+w(a^\dagger)^2].
\quad
\end{eqnarray}
Equivalently we can rewrite this as
\begin{eqnarray}
&&Z\left(|w|N+\bar w a^2,|w|N+w(a^\dagger)^2\right)
\nonumber\\
&&\quad
 =\left(1+2|w|f(2|w|,2|w|)\right) \{2|w|N+\bar w a^2+w(a^\dagger)^2\}
\nonumber\\
&&\qquad\qquad
+2 |w|^2 f(2|w|,2|w|)I
\nonumber\\
&&\quad
 ={e^{2|w|}-1\over 2|w|} \{2|w|N+\bar w a^2+w(a^\dagger)^2\}
\nonumber\\
&&\qquad\qquad
+{e^{2|w|}-(1+2|w|)\over 2} I.
\end{eqnarray}

\bigskip
\noindent
Numerous other results along quite similar lines can also be developed. Overall, this analysis provides a slightly different viewpoint on, and some possible extensions of, the usual squeezed-state formalism~\cite{Fisher, Schumaker, Truax, Shanta:1993, Nieto:1993, Nieto:1996}.

\section{Discussion}

While the general Baker--Campbell--Hausdorff formula for a free algebra is quite messy, 
various special cases where the commutator $[X,Y]$ is sufficiently simple that the iterates $L_X[X,Y]$ and $L_Y[X,Y]$, (and consequently, \emph{mutatis mutandi}, repeated iterations such as $L^n_X[X,Y]$ and $L^n_Y[X,Y]$),  are easy to calculate often lead to quite tractable closed-form algebraic expressions for $Z(X,Y) = \ln(e^X e^Y)$. We have illustrated such behaviour by considering the reasonably general but still tractable commutator
\begin{equation}
[X,Y]=uX+vY+cI,
\end{equation}
and extracting an exact analytic closed-form formula for the quantity $Z(X,Y) = \ln(e^X e^Y)$.  This result appears to be both new and non-trivial. To place this result in context, we have compared it with various other special cases already appearing in the literature.
Since this article first appeared online, the techniques have been adapted by Matone~\cite{Matone:2015a, Matone:2015b} to consider yet more general commutator algebras. 

\bigskip
\acknowledgments

This research was supported by the Marsden Fund, through a grant administered by the Royal Society of New Zealand. 

AVB was also supported by a Victoria University of Wellington Summer Scholarship.



\end{document}